\documentstyle[stwol,bigg10,psfig]{article}
\def\e{{\rm e}}
\def\del{\partial}
\def\half{{1\over2}}
\def\cosb{\cos\beta}
\def\sinb{\sin\beta}
\newcommand{\PRD}[3]{Phys. Rev. {\bf D{#1}} (19{#2}) {#3}}

\newcommand{\NPB}[3]{Nucl. Phys. {\bf B{#1}} (19{#2}) {#3}}

\newcommand{\PTP}[3]{Prog. Theor. Phys. {\bf {#1}} (19{#2}) {#3}}

\begin{document}
\title{ Profile of the Electroweak Bubble Wall \footnote{}}
\author{ K.~Funakubo }
\address{Department of Physics, Saga University, Saga 840, Japan}
\author{ A.~Kakuto, S.~Otsuki and F.~Toyoda }
\address{Department of Liberal Arts, Kinki University in Kyushu,
Iizuka 820, Japan}
\author{K.~Takenaga}
\address{Department of Physics, Kobe University, Kobe 657, Japan}

\twocolumn[\maketitle\abstracts{
We study the CP-violating bubble wall
by solving the coupled equations of motion derived from
the two-Higgs-doublet model for the moduli  $\rho_i$  and
phases $\theta_ i$ of the two scalars $(i=1,2)$ at the transition
temperature.  We find a solution that the CP-violating phase
$\theta\equiv\theta_1-\theta_2$  connecting the
CP-conserving vacua  strongly violates CP in the intermediate region while
the moduli largely deviate from the kink shape there.
We  estimate the chiral charge flux through the wall surface, which
may be efficient enough to produce
the baryon asymmetry by the sphaleron process.}]

\section{Introduction}
\footnotetext{Talk presented by F.~Toyoda at 28th International
Conference on High-energy Physics (ICHEP 96), Warsaw 1996. To be published in
the Proceedings.}
It is essential in any scenario of the electroweak baryogenesis \cite{reviewEB}
to know the
spacetime-varying profile of the CP violation  created at the electroweak
phase transition. In literatures,
some functional forms of the profile were assumed \cite{NKC,FKOTTa}.
They should be determined, however, by the dynamics of the gauge-Higgs
system, in which
the modulus and phase of the profile would be governed by some classical
equations of motion derived from the effective Higgs potential at the
transition temperature.

In a previous paper \cite{FKOTTb}, we presented a
solution in the two-Higgs-doublet model such that the CP-violating phase
$\theta$
spontaneously generated becomes as large as $O(1)$ around the wall while
it completely vanishes  in the broken and symmetric phase limits.
In the solution, the moduli $\rho_i$'s of the scalars were fixed to be
the kink shape. This solution, which we call as `solution with kink
ansatz', is interesting since it does yield an efficient
amount of the chiral charge flux through the wall surface,
which will be turned into the baryon density in the symmetric phase
region by the sphaleron transition. From the dynamical point of view, however,
such a large $\theta$ would affect $\rho_i$'s in turn.

In this article, we solve coupled equations of motion for $(\rho_i,\theta_i)$
derived from the two-Higgs-doublet model by imposing the discrete symmetry
$(\rho_1,\theta_1)\leftrightarrow(\rho_2,\theta_2)$. In this solution, which
we call as `solution without kink ansatz', $\rho_i$'s deviate largely
from the kink shape while $\theta$ remains to be of $O(1)$. We also estimate
the chiral charge flux due to this solution, which is
found also to be  sufficiently large.

\section{Equations of Motion}
At the temperature $T_C$ of the phase transition, the neutral
components of the  Higgs fields develop VEV's as the profile of the
bubble wall:
\begin{equation}
  \langle\Phi_i(x)\rangle
 =\left(\begin{array}{c} 0 \\
        {1\over{\sqrt2}}\rho_i(x)\e^{i\theta_i(x)}
        \end{array}\right),(i=1,2).
        \label{eq:VEV}
\end{equation}
Let the effective Higgs potential at $T_C$ be $V_{eff}(\rho_i,\theta_i)$.
Regarding the bubble wall as a static planar object, the equations of motion
are given by
\begin{eqnarray}
 {{d^2\rho_i(z)}\over{dz^2}}-\rho_i(z)\left({{d\theta_i(z)}\over{dz}}\right)^2
  -{{\del V_{eff}}\over{\del\rho_i}} &=& 0,
        \label{eq:rho-z}  \\
 {d\over{dz}}\left(\rho_i^2(z){{d\theta_i(z)}\over{dz}}\right)
  -{{\del V_{eff}}\over{\del\theta_i}} &=& 0,
        \label{eq:theta-z}
\end{eqnarray}
where $z$ is the coordinate perpendicular to the wall.
{}From the requirement that the gauge fields are pure-gauge type,
a constraint equation, ``sourcelessness condition", is added to them:
\begin{equation}
  \rho_1^2(z){{d\theta_1(z)}\over{dz}}+\rho_2^2(z){{d\theta_2(z)}\over{dz}}=0.
        \label{eq:soureless-z}
\end{equation}

As $V_{eff}$ at $T_C$, we postulate the same one examined in the previous
paper \cite{FKOTTb}:
\begin{eqnarray}
 V_{eff}(\rho_i,\theta_i)&=&
 \half m_1^2\rho_1^2+\half m_2^2\rho_2^2 + m_3^2\rho_1\rho_2\cos\theta
    \nonumber \\
&+&{{\lambda_1}\over8}\rho_1^4+{{\lambda_2}\over8}\rho_2^4   \nonumber  \\
&+&{{\lambda_3-\lambda_4}\over4}\rho_1^2\rho_2^2
 - {{\lambda_5}\over4}\rho_1^2\rho_2^2\cos2\theta \nonumber \\
&-& \half(\lambda_6\rho_1^2+\lambda_7\rho_2^2)\rho_1\rho_2\cos\theta
    \nonumber \\
&-& (A\rho_1^3+B\rho_1^2\rho_2\cos\theta \nonumber \\
&+& C\rho_1\rho_2^2\cos\theta + D\rho_2^3 ),\label{eq:geneVeff}
\end{eqnarray}
where $\theta\equiv\theta_1-\theta_2$. Here all the coefficients are
understood to include the radiative and the finite-temperature corrections,
and are assumed to be real so that CP is violated spontaneously.
The $\rho^3$ terms are expected to arise at finite temperatures so that
the moduli of VEV's in (\ref{eq:VEV}) take the kink shape for
$\theta=0$ as below.
\subsection{Kink ansatz}
In order for the kink shape moduli,
\begin{eqnarray}
      \rho_1(z)&=&{v\over 2}\cos\beta(1+\tanh(az)),\nonumber \\
      \rho_2(z)&=&{v\over 2}\sin\beta(1+\tanh(az)),\label{kink}
\end{eqnarray}
to be dynamically realized for $\theta(z)=0$, the parameters in
(\ref{eq:geneVeff})
have to satisfy certain set of relations. Here $v\cos\beta$ and $v\sin\beta$
are VEV's of $\Phi_1$ and $\Phi_2$ respectively, and $1/a$ is the common
wall width. As in the previous paper, it is convenient to use the following
parameters:
\begin{eqnarray}
  b&\equiv& -{{m_3^2}\over{4a^2\sinb\cosb}},   \nonumber\\
  c&\equiv& {{v^2}\over{32a^2}}(\lambda_1\cot^2\beta+\lambda_2\tan^2\beta
\nonumber \\
   &+&2(\bar\lambda_3-\lambda_5)) - {1\over{2\sin^2\beta\cos^2\beta}}
      \nonumber \\
   &=&  {{v^2}\over{8a^2}}(\lambda_6\cot\beta + \lambda_7\tan\beta),
\label{eq:def-bcd}\\
  d&\equiv& {{\lambda_5 v^2}\over{4a^2}}.   \nonumber   \\
  e&\equiv& {v\over{4a^2\sin^2\beta\cos^2\beta}}
             \BIG( A\cos^3\beta \nonumber \\
   &+& D\sin^3\beta-{{4a^2}\over v} \BIG) \nonumber\\
   &=& -{v\over{4a^2}}\left({B\over\sinb}+{C\over\cosb}\right) \nonumber
\end{eqnarray}
where $\bar\lambda_3\equiv\lambda_3-\lambda_4$,
and we have used the relations among parameters to rewrite $c$ and $e$.
The condition that  $(\rho_1,\rho_2)=(0,0)$ and $(\rho_1,\rho_2)=
(v\cos\beta,v\sin\beta)$ are  the local minima of $V_{eff}$ for
$\theta(z)=0$ leads to inequalities among the parameters:
\begin{equation}
    b>-1,\quad b-2e+3c>-1+(\bar\lambda_3-\lambda_5)v^2/4a^2.
\end{equation}
The equations of motion together with the ``sourcelessness condition"
are reduced to a single equation for $\theta$ with the kink ansatz
as examined in the previous paper.
\subsection{Discrete symmetry without kink ansatz}
When we do not impose the kink ansatz, we have many free parameters. To reduce
the number of them, we require that $V_{eff}$ is symmetric under
$(\rho_1,\theta_1)\leftrightarrow(\rho_2,\theta_2)$ and that this discrete
symmetry is not spontaneously broken.
This means that the parameters in (\ref{eq:geneVeff}) should satisfy
\begin{eqnarray}
 m_1^2&=&m_2^2\equiv m^2, \nonumber\\
 \lambda_1&=&\lambda_2,\quad \lambda_6=\lambda_7,\quad  \label{eq:sym-para}
 A=D,\quad B= C,\quad
\end{eqnarray}
and that $\tan\beta=1$. For
\begin{eqnarray}
 \rho_1(z)&=&\rho_2(z)\equiv\rho(z)/{\sqrt{2}},\nonumber\\
 \theta_1(z)&=&-\theta_2(z)\equiv\theta(z)/2,
\end{eqnarray}
we have two coupled equations:
\begin{eqnarray}
 {{d^2\rho(z)}\over{dz^2}}
 -{1\over4}\rho(z)\left({{d\theta(z)}\over{dz}}\right)^2
  -{{\del V_{eff}}\over{\del\rho}} &=& 0,     \label{eq:sym-rho-z}  \\
 {1\over4}{d\over{dz}}\left(\rho^2(z){{d\theta(z)}\over{dz}}\right)
  -{{\del V_{eff}}\over{\del\theta}} &=& 0,   \label{eq:sym-theta-z}
\end{eqnarray}
while ``sourcelessness condition" is automatically satisfied.
Under the discrete symmetry, there remains only one
free parameter in $V_{eff}$ once $b$, $c$, $d$ and $e$ are given.
We take $\lambda_1$ as the free parameter.
Since $\tan\beta=1$, the parameters are given by
\begin{eqnarray}
 m_3^2 &=& -2 a^2 b,\quad
 m^2=4a^2 - m_3^2,\quad
 \lambda_5={{4a^2}\over{v^2}} d,\nonumber\\
 \bar\lambda_3&=&\lambda_5 + {{16a^2}\over{v^2}}(c+2) - \lambda_1,\nonumber \\
 \lambda_6 &=& {{\lambda_1+\bar\lambda_3-\lambda_5}\over4}
          -{{8a^2}\over{v^2}},\nonumber\\
 A&=&{{\sqrt{2}a^2}\over v}(e+4),\quad
 B=-A + {{4\sqrt{2}a^2}\over v}.
\end{eqnarray}

\section{Solutions with and without Kink Ansatz}
\subsection{Solution with kink ansatz}
The parameter set for this type of solution is constrained from the asymptotic
behaviors of $\theta$ in the background of the kink-shape $\rho$ in (2.6).
One of the solutions presented in the previous paper \cite{FKOTTb}
is obtained for the set
$(b,c,d,e)=(3,12.2 ,-2,12.2)$, which is an example satisfying the
constraints \cite{FKOTTb}.
The boundary conditions for this solution is $\theta_b\equiv
\theta(+\infty)=0$($\theta_s\equiv\theta(-\infty)=0$) in the broken(symmetric)
 phase limit.  The profile of the bubble wall of this solution is given
in Fig.1. Note that the CP-violating imaginary part is large just around
the wall.

\subsection{Solution without kink ansatz}
We impose the boundary conditions $\rho_b\equiv\rho(+\infty)=v$
($\rho_s\equiv\rho(-\infty)=0$) together with $\theta_b=0(\theta_s=0)$
in the broken(symmetric) phase limit.
We take the same values of $(b,c,d,e)$ for comparison, though they do
not need to be constrained. For definiteness, we set
$(v,a)=(246,10)$ in the unit of GeV. The profile of
the bubble wall is shown in Fig.2. The real part of the wall is
drastically altered from the kink shape. The imaginary part of the wall is
much larger than that of the solution with the kink ansatz.

\section{Concluding Remarks}
We have investigated the CP-violating profile of the wall by solving
the equations of motion from the two-Higgs-doublet model.
The most remarkable feature may be that,
while CP is conserved in the broken and symmetric phase limits,
the CP-violating phase $\theta$
becomes $O(1)$ in the intermediate region.
The model could  avoid the light scalars due to the Georgi-Pais
theorem \cite{GP}. We give some comments.
\par\noindent
(1) The baryon asymmetry of the universe through the sphaleron
process along the charge transport scenario is roughly given by \cite{FKOTTc}
$$
\frac{\rho_B}{s}\simeq {10^{-7}\over T^2}\,\,{F_Q\over u}\,\,\tau,
$$
where the entropy density is given by $s=2\pi^2g_*T^3/45 $ with
$g_*\simeq100$, $u$ is the wall velocity, $\tau$ is the transport
time within which the scattered fermions are captured by the wall
and $F_Q$ is the chiral charge flux through the CP-violating wall.
Fig.3 gives the contour plot of $F_Q$
due to the solution without kink ansatz for various choices of $m/T$ and
$a/T$, where $m$ is the relevant fermion mass
\footnote{For a numerical method how to obtain the chiral transmission and
reflection coefficients for such solutions, see \cite{FKOTa}.}.
This should be compared with the corresponding one due to the solution with
kink ansatz given in the previous paper. For the both solutions, $F_Q$
may be said to be large enough because of the large $\theta$'s
around the wall surface. This would be
important since there are many possible mechanisms to diminish the net baryon
asymmetry in the baryogenesis
scenarios, whether spontaneous or charge transport \cite{reviewEB}.
\par\noindent
(2) It may be interesting to see how the deviation of $\rho$ from the
kink shape affects the  energy density of the wall
per unit area, which is given as
\begin{eqnarray}
 {\cal E}&=&\int_{-\infty}^\infty dz {\BIG\{}
  \half\sum_{i=1,2}\left[\left({{d\rho_i}\over{dz}}\right)^2
    +\rho_i^2 \left({{d\theta_i}\over{dz}}\right)^2 \right]\nonumber\\
   &+& V_{eff}(\rho_1,\rho_2,\theta) {\BIG\}}.     \nonumber
\end{eqnarray}
For the trivial solution $\theta\equiv 0$ with the kink ansatz,
$\left.{\cal E}\right|_{\theta=0}=av^2/3$.
The energy densities of the above solutions are
$$
   \Delta{\cal E}\equiv{\cal E} - \left.{\cal E}\right|_{\theta=0} =
\left\{\begin{array}{l}
                 - 2.056\times 10^{-3}\,av^2\sin^2\beta\cos^2\beta, \\
                 - 3.473\times 10^{-2}\,av^2\sin^2\beta\cos^2\beta,
\end{array}\right. \nonumber
$$
where the
upper value is obtained with kink ansatz and the lower one
without kink ansatz, and
we fix $\sin^2\beta\cos^2\beta = 1/4$ because it is required for
the latter case by the discrete symmetry.
The enhancement factor ${\rm exp}\left( -{4\pi R_C^2\Delta{\cal E}}/{T_C}
\right)$ to form the respective bubble over that with the trivial solution
is
$$
 \exp\left( -{{4\pi R_C^2\Delta{\cal E}}\over{T_C}}\right) =
\left\{
\begin{array}{ll}
 1.25 &(\mbox{kink}),\\
 43.7 &(\mbox{nonkink}),
\end{array}\right. \nonumber
$$
where the radius of the critical bubble $R_C$ is approximately
given by $\sqrt{3F_C/(4\pi av^2)}$ with $F_C$ being the
free energy of the critical bubble, and $F_C=145T$ is taken as
various authors estimate $F_C\simeq(145\sim 160)T$.
\par\noindent
(3) Since $V_{eff}$ is symmetric under the exchange of
$\theta \leftrightarrow -\theta$,
it is expected that the bubble with $\theta>0$ and that with $\theta<0$
are created with the equal probability, so that the chiral charge flux or
the net baryon number density would be averaged to vanish. An explicit
CP breaking to violate the symmetry could avoid this difficulty
as pointed out by Comelli $et\,al.$ \cite{Comelli}. In another
article \cite{FKOTb} we give an estimate that, even if the
explicit breaking at the phase transition is as small as the
Kobayashi-Maskawa scheme, the relative enhancement factor
between the two kinds of bubbles is able to amount to $O(10)$ when
the explicit breaking is taken into account with kink ansatz.

\vspace{4cm}
\centerline{\ }

\psfig{figure=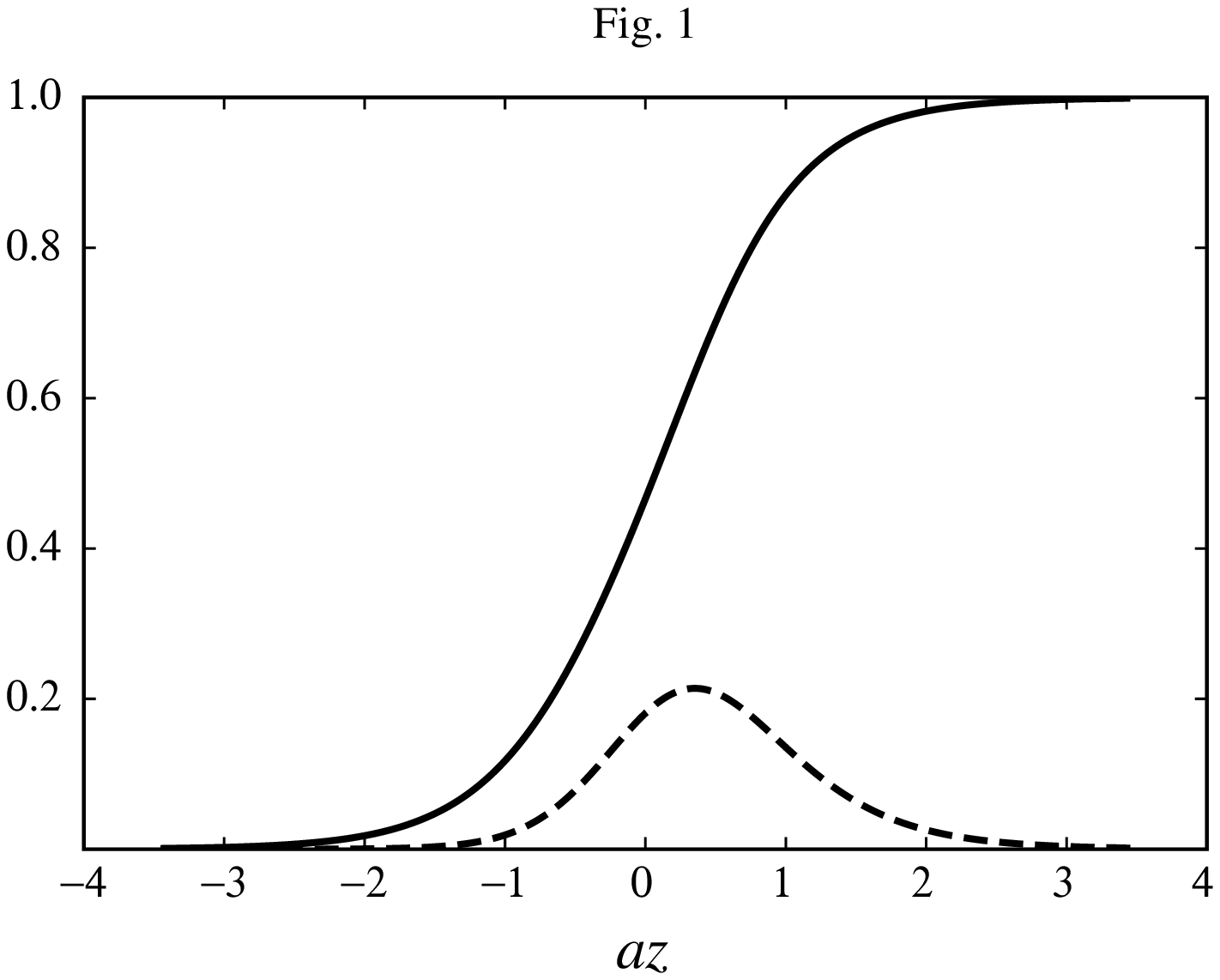,height=5cm}
\noindent The profile of the bubble wall corresponding to the solution
with kink ansatz.The solid line represents the real part of the profile.
The dashed line is the imaginary part.
\psfig{figure=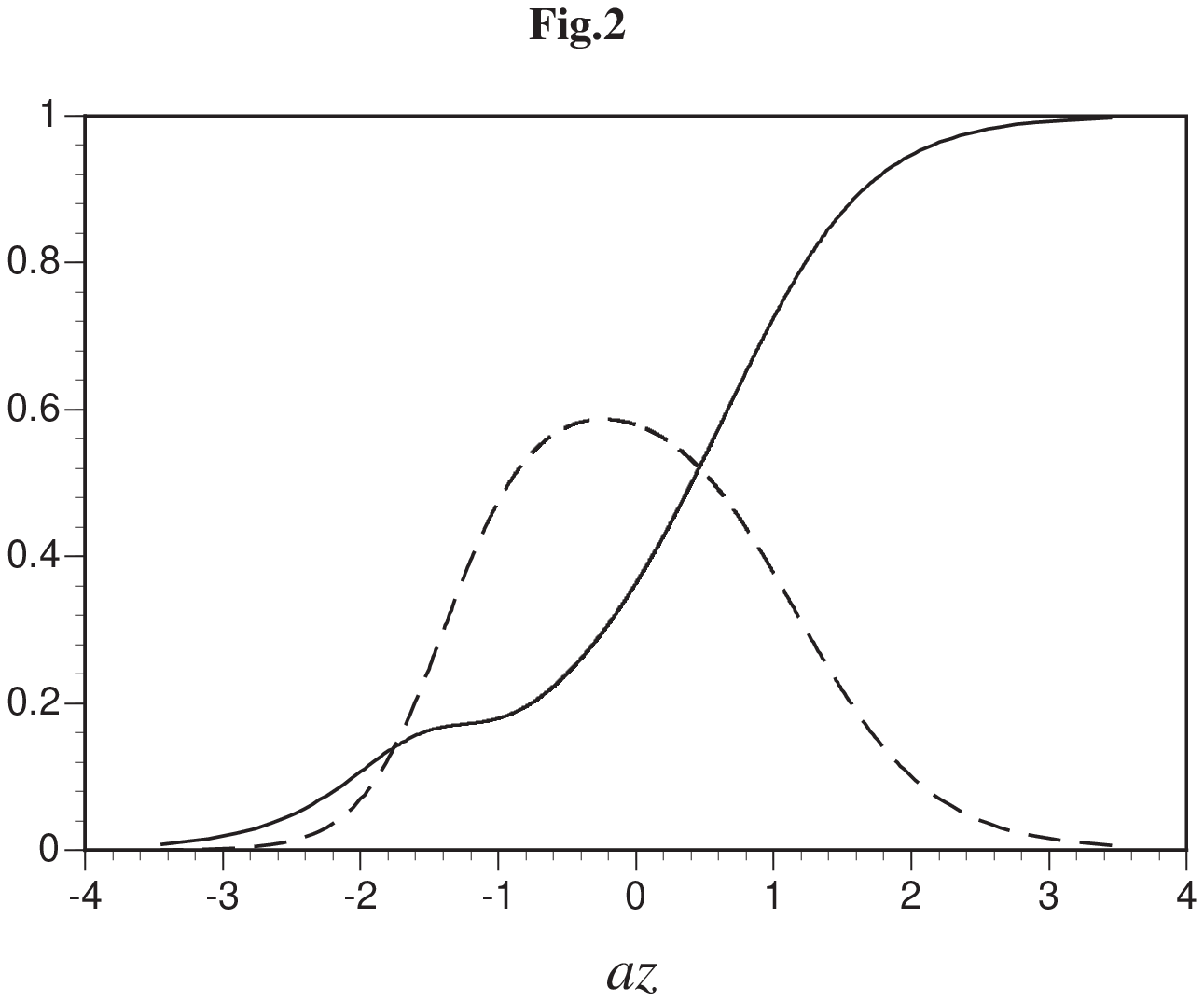,height=5cm}
\par\noindent The profile of the bubble wall without kink ansatz.
The solid and dashed lines are real and imaginary parts of the profile,
respectively.
\psfig{figure=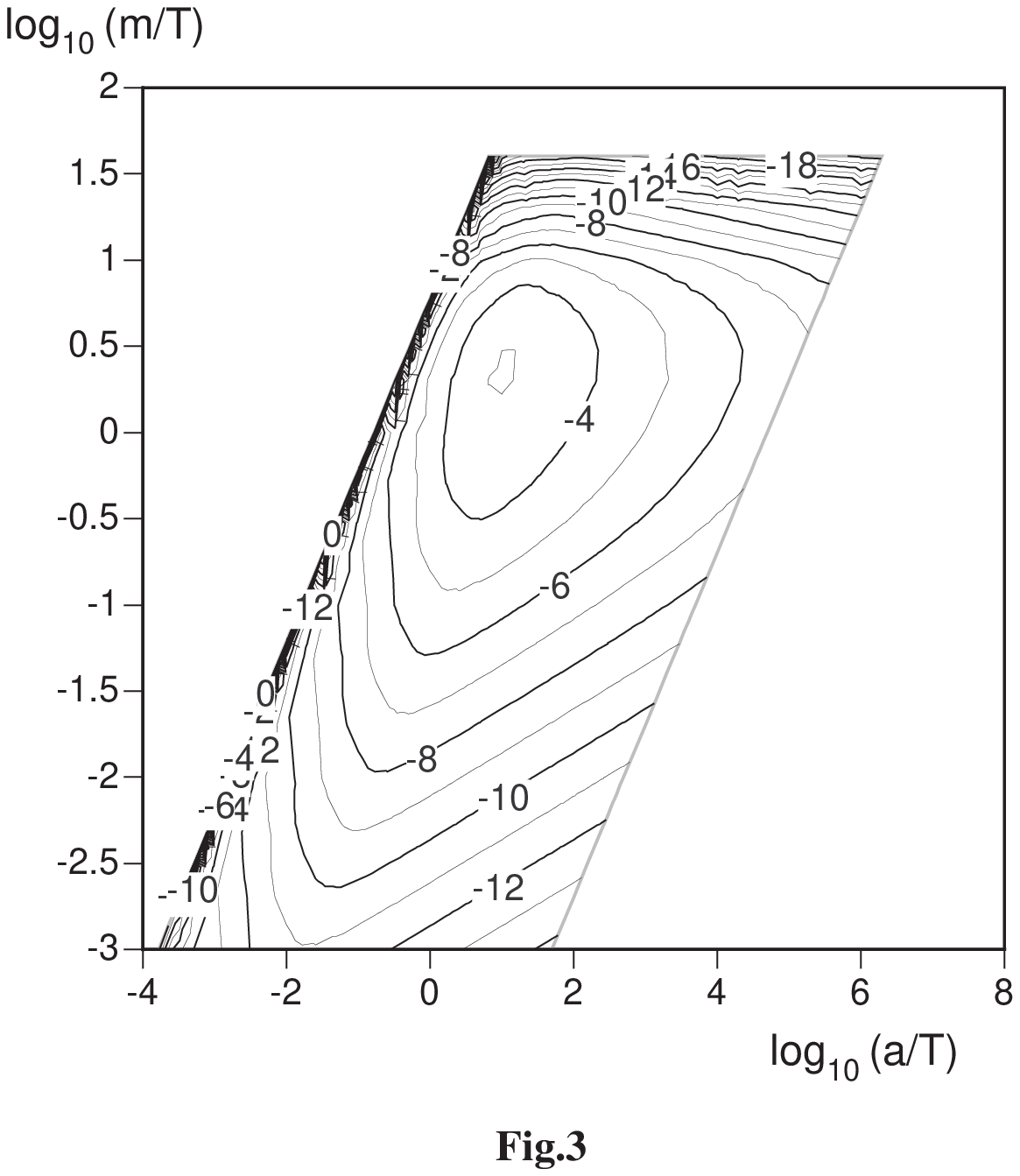,height=5cm}
\par\noindent Contour plot of $F_Q/(uT^3(Q_L-Q_R))$, for $u=0.1$ and
$T=100$GeV, where $u$ is the wall velocity, $Q_{L(R)}$ is the chiral
charge of the left(right) handed fermion with the mass $m$, and $1/a$ is
the wall width.
\end{document}